\documentclass[10pt,aps,pra,notitlepage,nofootinbib,superscriptaddress,noeprint]{revtex4-2}
\usepackage[english]{babel}
\usepackage[T1]{fontenc}
\usepackage{amssymb}
\usepackage{xcolor}
\usepackage{graphicx}
\graphicspath{{Figures/}{./}}
\usepackage{mathtools}
\usepackage{physics}
\usepackage{microtype}
\usepackage{dsfont}
\usepackage[breaklinks=true,colorlinks=true,linkcolor=teal,urlcolor=teal,citecolor=teal]{hyperref}
\usepackage{orcidlink}

\begin{document}

\title{Pulsed single-photon magnetometry with a \texorpdfstring{$\Lambda$}{Lambda}-type three-level system: near-optimal frequency-resolved photon counting}

\author{Seyed Mostafa Moniri\orcidlink{0000-0003-1738-4429}}
\email{s.m.moniri@gmail.com}
\affiliation{Basic Sciences Group, Golpayegan College of Engineering, Isfahan University of Technology, Golpayegan 87717-67498, Iran}
\author{Elnaz Darsheshdar\orcidlink{0000-0001-6341-7151}}
\email{darsheshdare@gmail.com}
\affiliation{Department of Physics, University of Warwick, Coventry, CV4 7AL, United Kingdom}
\affiliation{Institute for Physical Research, Ashtarak-2, 0203, Armenia}
\author{Mikayel Khanbekyan\orcidlink{0009-0009-2026-1164}}
\email{khanbekyan@gmail.com}
\affiliation{Institute for Physical Research, Ashtarak-2, 0203, Armenia}

\date{\today}

\begin{abstract}
We investigate pulsed single-photon magnetometry with a Zeeman-sensitive $\Lambda$-type three-level system driven by a classical control field. We derive the asymptotic output state and decompose its quantum Fisher information into photon-loss, spectral-intensity, and spectral-phase contributions. Environmental coupling reshapes the scattering response and can increase magnetic-field information. At critical coupling, real-frequency zeros of the scattering amplitude redistribute information toward measurable spectral intensity, allowing frequency-resolved photon counting to capture nearly all of the magnetic-field information encoded in the output state when the zeros lie within the pulse bandwidth. For long Gaussian pulses with a smooth, nonzero central-frequency scattering amplitude, the additional spectral-intensity contribution and residual spectral-phase information gap decrease as $T^{-2}$ or faster.
\end{abstract}

\maketitle

Pulsed quantum-light spectroscopy studies how information about a quantum matter system can be extracted by measuring a light pulse after it has interacted with the system. Previous studies used a single-photon pulse interacting with a two-level system (TLS) to estimate the light-matter coupling constant $\Gamma$, which is related to the electric dipole moment~\cite{Albarelli2023,Darsheshdar2024}. In such systems, the parameter dependence appears in the output state through photon loss, scattering-induced phase shifts, and distortions of the pulse temporal profile. This suggests a broader question: beyond coupling-constant estimation, can pulsed single-photon scattering be used to extract information about external fields?

In this paper, we address this question using a Zeeman-sensitive $\Lambda$-type three-level system in an external magnetic field $B$. The system is probed by a single-photon pulse on the $\ket{g}\leftrightarrow\ket{e}$ transition, while a classical control field drives the $\ket{s}\leftrightarrow\ket{e}$ transition. Such $\Lambda$-type systems are central to electromagnetically induced transparency (EIT), dark-state physics, and Raman-resonant coherent control \cite{Fleischhauer2000DSP,Fleischhauer2005EIT,Vitanov2017STIRAP}. Related EIT phenomena have also been studied in magnetically sensitive alkali-vapor systems and in the dark-state-polariton description of coherently driven multilevel media \cite{SargsyanPapoyan2024,ZimmerUnanyan2008}.
Single-photon interactions with classically driven $\Lambda$ systems have also been explored in cavity and waveguide settings, including controlled photon absorption, coherent single-photon scattering, and controlled generation and shaping of single-photon wave packets~\cite{Dilley2012,Yadav2025,Khanbekyan2017}. Related classically driven three-level emitter schemes have also been proposed for time-bin entangled photon-pair generation \cite{Khanbekyan2018TimeBin}. The sensitivity of EIT resonances to Zeeman shifts has motivated their use in atomic magnetometry~\cite{Lee1998EIT}. More broadly, resonance-enhanced optical responses have been exploited for sensing in cavity and scattering settings, including exceptional-point-assisted discrimination and magnetic-field-sensitive light scattering~\cite{Khanbekyan2022,Mkrtchian2026}.

Compared with a two-level system, in which the magnetic field shifts a single optical resonance, the $\Lambda$ configuration introduces a control-tunable scattering pathway and a two-photon (Raman) detuning governed by the differential Zeeman shift of the two lower states. The magnetic field therefore changes both the one-photon and two-photon detunings and modifies the interference between the excitation pathways. As a result, magnetic-field information can be encoded in both the photon-loss statistics and the spectral properties of the single-photon output state.

Our central objective is to identify where the magnetic-field information is stored in the output state and to determine whether frequency-resolved photon counting can extract information beyond a photon-loss measurement. 
We derive the asymptotic output state for an arbitrary incident single-photon pulse and separate its quantum Fisher information (QFI) into classical and quantum contributions. 
The classical contribution arises from photon-loss statistics and is associated with the vacuum-versus-one-photon probabilities, while the quantum contribution is carried by the conditional single-photon output state and is further
resolved into spectral-intensity and spectral-phase terms.
We then specialize the results to Gaussian pulses and compare a photon-loss measurement with frequency-resolved photon counting. This analysis identifies the conditions under which frequency-resolved photon counting closely approaches the output-state QFI and clarifies the roles of environmental coupling and real-frequency zeros of the scattering amplitude.

\section{Model and Zeeman encoding}\label{sec:model}
\subsection{Zeeman shifts and magnetic-field-dependent detunings}
\begin{figure}[t]
	\centering
	\includegraphics[width=0.8\linewidth]{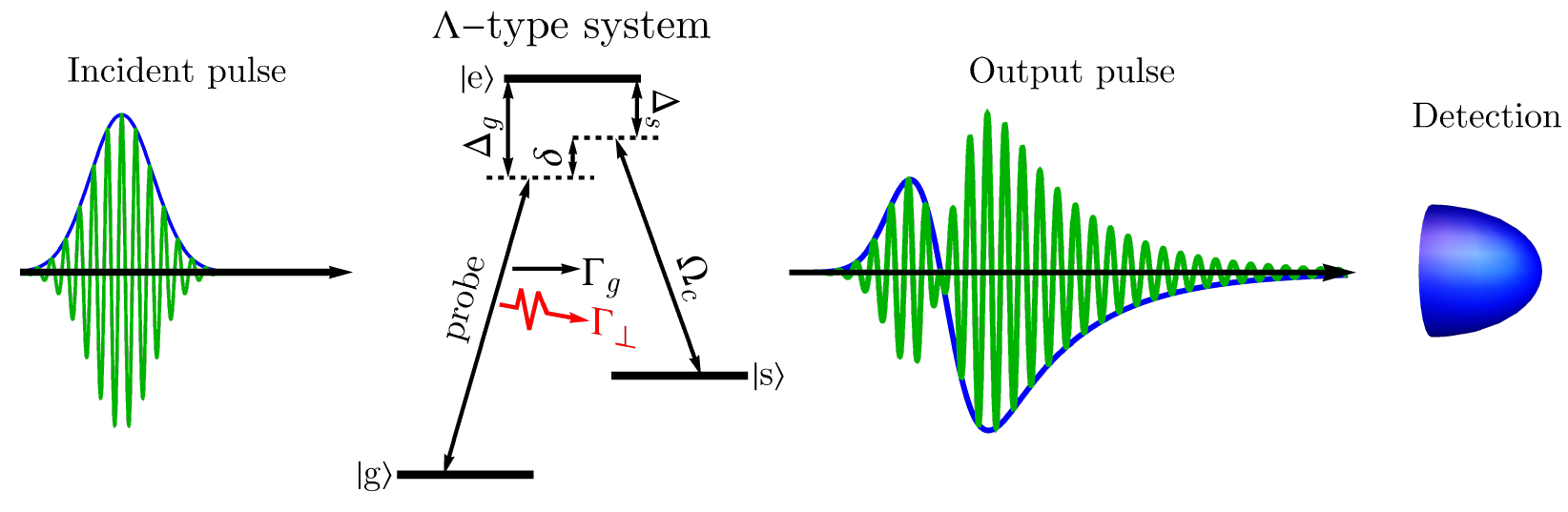}
	\caption{Schematic of pulsed single-photon magnetometry with a Zeeman-sensitive $\Lambda$-type three-level system. An incident single-photon pulse couples the $\ket{g}\leftrightarrow\ket{e}$ transition at rate $\Gamma_g$, while a classical control field with Rabi frequency $\Omega_c$ drives the $\ket{s}\leftrightarrow\ket{e}$ transition. The magnetic field modifies the one-photon detuning $\Delta_g$ and the two-photon detuning $\delta$. Emission into environmental modes is described by $\Gamma_\perp$.}
    \label{fig:schematic}
\end{figure}

We consider a $\Lambda$-type three-level system with states $|g\rangle, |s\rangle, |e\rangle$. A schematic of the model is shown in Fig.~\ref{fig:schematic}. 
A single-photon pulse probes the transition $|g\rangle \leftrightarrow |e\rangle$ while a classical control field drives the transition $|s\rangle \leftrightarrow |e\rangle$. In the linear Zeeman regime, an external magnetic field $B$ shifts the atomic levels by
\begin{equation}
\Delta E_j^{(Z)}=\mu_B \mathfrak{g}_j m_j B,
\end{equation}
where $\mu_B$ is the Bohr magneton, $\mathfrak{g}_j$ is the Land{\'e} factor, and $m_j$ is the magnetic quantum number of the state $|j\rangle$. Therefore, the transition frequencies become magnetic-field dependent:
\begin{equation}
\begin{aligned}
\omega_{eg}(B)=\omega_{eg}^0+\frac{\mu_B B}{\hbar}\left(\mathfrak{g}_e m_e-\mathfrak{g}_g m_g\right),\\
\omega_{es}(B) =\omega_{es}^0+\frac{\mu_B B}{\hbar}\left(\mathfrak{g}_e m_e-\mathfrak{g}_s m_s\right).
\end{aligned}
\label{frequencyshift}
\end{equation}
The corresponding one-photon detunings are
\begin{equation}
\Delta_g(B) =\omega_p-\omega_{eg}(B),\qquad
\Delta_s(B) =\omega_c-\omega_{es}(B),
\label{onephotondetuning}
\end{equation}
where $\omega_p$ is the central frequency of the single-photon pulse and $\omega_c$ is the frequency of the classical control field. We define the two-photon detuning as
\begin{equation}
\delta(B) =\Delta_g(B) -\Delta_s(B).
\label{twophotondetuning}
\end{equation}
The condition $\delta=0$ corresponds to two-photon resonance, where the system can support a dark state and exhibit EIT. The magnetic-field dependence enters the scattering dynamics through the
detunings $\Delta_g$ and $\delta$. For fixed $\omega_p$ and $\omega_c$, Eqs.~\eqref{frequencyshift}-\eqref{twophotondetuning} give
\begin{equation}
\begin{aligned}
&\kappa_g\equiv\partial_B \Delta_g(B) = -\frac{\mu_B}{\hbar}(\mathfrak{g}_e m_e-\mathfrak{g}_g m_g),\\
&\kappa_\delta\equiv\partial_B \delta(B) = \frac{\mu_B}{\hbar}(\mathfrak{g}_g m_g-\mathfrak{g}_s m_s).
\end{aligned}
\end{equation}
Around a reference field $B_0$, the magnetic-field dependence of the detunings can be written as
\begin{equation}
\Delta_g(B)=\Delta_{g,0}+\kappa_g(B-B_0),\qquad
\delta(B)=\delta_0+\kappa_\delta(B-B_0),
\label{eq:B_detuning_dependence}
\end{equation}
where $\Delta_{g,0}=\Delta_g(B_0)$ and $\delta_0=\delta(B_0)$. Thus, the two-photon detuning is sensitive to the differential Zeeman shift of the two lower states, while the one-photon detuning generally changes at the same time. Near the two-photon resonance condition $\delta\simeq 0$, even a small change in $B$ can therefore strongly modify the single-photon scattering response.

%\subsection{Choice of Zeeman sublevels, field geometry, and polarizations}
We take $\ket{g}$, $\ket{s}$, and $\ket{e}$ to denote three selected Zeeman sublevels
\begin{equation}
\ket{g}=\ket{J_g,m_g},\qquad
\ket{s}=\ket{J_s,m_s},\qquad
\ket{e}=\ket{J_e,m_e},
\end{equation}
rather than complete fine-structure manifolds. The magnetic field defines the quantization axis, which we take along $z$. The single-photon pulse and classical control field obey the electric-dipole selection rules for the two transitions. The  $\pi$ polarization drives $\Delta m=0$, while $\sigma^{+}$ and $\sigma^{-}$ polarizations drive $\Delta m=+1$ and $\Delta m=-1$, respectively.
The assumed unidirectional propagation is compatible with a definite polarization. The effective three-level description is valid when the selected transitions are well resolved in frequency and polarization, so that coupling to other Zeeman sublevels is negligible. This requires a well-defined quantization axis, controlled pulse and control-field polarizations, and Zeeman splittings larger than the incident pulse bandwidth and all relevant driven-system frequency scales, including the natural linewidths and the scale set by $\lvert\Omega_c\rvert$. Under these conditions, the neglected Zeeman sublevels give only small perturbative corrections.

\subsection{Interaction with the pulse and environment}
\label{sec:interactions}
In a rotating frame, the Hamiltonian of the three-level system is
\begin{equation}
H_{\mathrm{A}}= -\hbar \Delta_g|e\rangle\langle e|
-\hbar \delta|s\rangle\langle s| .
\label{H_atom}
\end{equation}
In what follows, we write $\Delta_g$ and $\delta$ for $\Delta_g(B)$ and $\delta(B)$, respectively, with their magnetic-field dependence given by Eq.~\eqref{eq:B_detuning_dependence}. We denote the pulse modes by the $P$ subsystem. The single-photon pulse couples to the $|g\rangle \leftrightarrow |e\rangle$ transition. We describe the pulse field by a continuum of quantum white-noise operators $a(t)$ within the broadband Markovian input-output formalism~\cite{Gardiner2004,Walls2008,Carmichael1993}. Throughout this work, an unsubscripted $\omega$ denotes the angular-frequency offset from the central frequency $\omega_p$ of the single-photon pulse, so that the corresponding absolute optical frequency is $\omega_p+\omega$. The Markov approximation assumes that the spectral density is approximately flat over the bandwidth of the incident single-photon pulse, $J_g(\omega_p+\omega)\simeq J_g(\omega_p)$, so that the coupling to the pulse modes is characterized by a constant rate $\Gamma_g$. The corresponding interaction Hamiltonian is
\begin{equation}
H_{AP}(t)= -i\hbar \sqrt{\Gamma_g}\left[a(t)\sigma_{eg}-a^\dagger(t)\sigma_{ge}\right],
\label{H_photon}
\end{equation}
where $\sigma_{eg}=|e\rangle\langle g|$, $\sigma_{ge}=|g\rangle\langle e|$, and $a(t)=\int_{-\infty}^{\infty}{d\omega}/{\sqrt{2\pi}} \,\, a(\omega)e^{-i\omega t}$. Here, $a(\omega)$ annihilates a photon in the pulse mode with frequency offset $\omega$. The bosonic operators satisfy the continuum commutation relation $[a(\omega),a^\dagger(\omega')]=\delta(\omega-\omega')$.

In addition to the pulse modes, the three-level system couples to other electromagnetic modes, including other spatial and polarization modes. We denote these environmental modes by the $E$ subsystem and describe them by an independent continuum of quantum white-noise operators $b(t)$, using the same approximation. The corresponding interaction Hamiltonian is
\begin{equation}
H_{AE}(t)= -i\hbar \sqrt{\Gamma_\perp} \left[b(t)\sigma_{eg}-b^\dagger(t)\sigma_{ge}\right],
\label{H_environment}
\end{equation}
where $\Gamma_\perp$ is the environmental coupling rate. Analogously, the environmental frequency-domain operators satisfy $[b(\omega),b^\dagger(\omega')]=\delta(\omega-\omega')$. The corresponding time-domain field operators obey the white-noise commutation relations, e.g.
\begin{equation}
[a(t),a^\dagger(t')]=\delta(t-t'),\qquad
[b(t),b^\dagger(t')]=\delta(t-t'),
\end{equation}
with the two continua taken to be mutually independent.

The classical control field drives the $|s\rangle \leftrightarrow |e\rangle$ transition,
\begin{equation}
H_C=\frac{\hbar}{2}\left(\Omega_c \sigma_{es}+ \Omega^*_c\sigma_{se}\right),
\label{H_classical}
\end{equation}
where $\sigma_{es}=|e\rangle\langle s|$ and $\sigma_{se}=|s\rangle\langle e|$. We treat the control as a strong classical field with Rabi frequency $\Omega_c$ taken to be constant over the scattering interval. Its relative quantum fluctuations and depletion due to the interaction with a single photon are neglected. The control therefore enters the dynamics only through $\Omega_c$. The total effective Hamiltonian is
\begin{equation}
\begin{aligned}
H(t)=&-\hbar \Delta_g|e\rangle\langle e|-\hbar \delta|s\rangle\langle s|+\frac{\hbar}{2}\left(\Omega_c \sigma_{es}+ \Omega^*_c\sigma_{se}\right)\\
&-i\hbar\left[\sqrt{\Gamma_g}a(t)\sigma_{eg}+\sqrt{\Gamma_\perp}b(t)\sigma_{eg}-\mathrm{h.c.}\right].
\end{aligned}
\label{H_total}
\end{equation}

We also neglect spontaneous branching from $\ket{e}$ to $\ket{s}$ at rate $\Gamma_s$ and dephasing of the lower-state coherence at rate $\gamma_{gs}$. Thus, $\Gamma_s=\gamma_{gs}=0$ in the ideal model. The state $\ket{s}$ is populated only coherently through $\Omega_c$, while $\Gamma_\perp$ accounts for emission into environmental modes. The conditions under which these approximations are valid, together with a possible experimental implementation, are discussed in Sec.~\ref{sec:implementation}.

\section{Single-photon scattering and metrological information}\label{sec:scattering}
\subsection{Time-domain scattering dynamics}
\label{sec:time_domain}
We now formulate the single-photon scattering dynamics of the Zeeman-sensitive $\Lambda$-type system. We consider the initial state $|\Psi(t_0)\rangle=|g\rangle|1_\xi\rangle_P|0\rangle_E$, where the three-level system is initially in the ground state $|g\rangle$, the environment is in vacuum, and the incoming single-photon pulse is
\begin{equation}
|1_\xi\rangle_P=\int_{-\infty}^{\infty} d\tau\,\xi(\tau)a^\dagger(\tau)|0\rangle_P ,
\end{equation}
where $\xi(t)$ is the temporal profile of the pulse, normalized as $\int_{-\infty}^{\infty}d\tau\,|\xi(\tau)|^2=1$. Under the assumptions of Sec.~\ref{sec:model}, the Hamiltonian conserves the total excitation number. Explicitly,
\begin{equation}
\hat N=\ket{e}\!\bra{e}+\ket{s}\!\bra{s}
+\int d\omega\,a^\dagger(\omega)a(\omega)
+\int d\omega\,b^\dagger(\omega)b(\omega),
\qquad [H(t),\hat N]=0.
\label{eq:excitation_number}
\end{equation}
The classical control field only interconverts the two atomic states $\ket{e}$ and $\ket{s}$ and therefore does not change the total excitation number. The evolution consequently remains in the single-excitation subspace, and the joint state can be written as
\begin{equation}
|\Psi(t)\rangle=\psi_e(t)|e\rangle|0\rangle_P|0\rangle_E+\psi_s(t)|s\rangle|0\rangle_P|0\rangle_E +|g\rangle\left[|\psi_g^P(t)\rangle|0\rangle_E+ |0\rangle_P|\psi_g^E(t)\rangle\right],
\label{eq:single_excitation_ansatz}
\end{equation}
where
\begin{align}
|\psi_g^P(t)\rangle=&\int_{-\infty}^{\infty} d\tau\,\psi_g^P(t,\tau)a^\dagger(\tau)|0\rangle_P ,\\
|\psi_g^E(t)\rangle=&\int_{-\infty}^{\infty} d\tau\,\psi_g^E(t,\tau)b^\dagger(\tau)|0\rangle_E .
\end{align}
These are the unnormalized single-photon states in the pulse and environment modes, respectively. Substituting Eq.~\eqref{eq:single_excitation_ansatz} into the Schr{\"o}dinger equation gives coupled equations for the internal amplitudes $\psi_e(t)$ and $\psi_s(t)$.  For the initial conditions $\psi_e(t_0)=\psi_s(t_0)=0$, with $t_0$ taken sufficiently early that the incident pulse is negligible, their solution can be written as
\begin{equation}
\begin{pmatrix}
\psi_e(t)\\
\psi_s(t)
\end{pmatrix}
=-\sqrt{\Gamma_g}\int_{t_0}^{t}dt'\,e^{M(t-t')}
\begin{pmatrix}
1\\
0
\end{pmatrix}
\xi(t'),
\label{eq:time_solution}
\end{equation}
where the coupled amplitude equations and the evolution matrix $M$ are given in Appendix~\ref{app:time_domain_solution}.
The unnormalized single-photon state in the pulse modes is determined by the input-output relation
\begin{equation} \label{psigp}
|\psi_g^P(t)\rangle=\int_{-\infty}^{\infty} d\tau\,\left[\xi(\tau)+\sqrt{\Gamma_g}\Theta(t-\tau)\psi_e(\tau)\right] a^\dagger(\tau)|0\rangle_P ,
\end{equation}
while the corresponding single-photon state in the environment modes is
\begin{equation} \label{psige}
|\psi_g^E(t)\rangle=\sqrt{\Gamma_\perp}\int_{-\infty}^{\infty} d\tau\,\Theta(t-\tau)\psi_e(\tau) b^\dagger(\tau)|0\rangle_E .
\end{equation}

The pulse output therefore results from the coherent interference between the incident photon and the field re-emitted into the pulse modes, whereas emission at rate $\Gamma_\perp$ transfers the excitation to the environment. These contributions determine the asymptotic joint state. In the following subsection, we obtain the corresponding frequency-domain scattering solution.

\subsection{Asymptotic frequency-domain scattering solution}\label{sec:freq_domain}

At asymptotic times, after the incident pulse has left the interaction region and the atomic transients have decayed, the pulse output can be described in the frequency domain. Since the control field is constant, the coupled equations have time-independent coefficients and can be solved in the frequency domain. We use the Fourier convention
\begin{equation}
\tilde f(\omega)=\int_{-\infty}^{\infty}\frac{dt} {\sqrt{2\pi}} \,f(t)e^{i\omega t}, \qquad f(t)=\int_{-\infty}^{\infty}\frac{d\omega}{\sqrt{2\pi}} \tilde f(\omega)e^{-i\omega t}. 
\label{FourierConvention}
\end{equation}
The unnormalized single-photon state in the pulse modes can then be written in the frequency domain as
\begin{equation}
|\psi_g^P(\infty)\rangle=\int_{-\infty}^{\infty}d\omega\,\tilde{\xi}(\omega)t_B(\omega) a^\dagger(\omega)|0\rangle_P ,
\label{psioutfreq}
\end{equation}
where $\tilde{\xi}(\omega)$ denotes the spectral amplitude of the incident pulse, and the frequency-dependent scattering amplitude is
\begin{equation}
t_B(\omega)=1-\Gamma_g\chi(\omega;B),
\label{eq:scattering_amplitude}
\end{equation}
and
\begin{equation} 
\chi(\omega;B) = \frac{\omega+\delta} { \left[\frac{\Gamma_g+\Gamma_\perp}{2}-i(\omega+\Delta_g)\right](\omega+\delta) +i\frac{|\Omega_c|^2}{4} }
\end{equation}
is the frequency-domain atomic response function. The factor $\omega+\delta$ reflects the contribution of the metastable-state pathway, while the control field enters through the term $|\Omega_c|^2/4$ in the denominator. The corresponding single-photon state in the environment modes is
\begin{equation} \label{psiEfreq}
|\psi_g^E(\infty)\rangle = - \sqrt{\Gamma_g\Gamma_\perp}\int_{-\infty}^{\infty} d\omega\, \tilde{\xi}(\omega) \chi(\omega;B) b^\dagger(\omega)|0\rangle_E .
\end{equation} 
We define the unnormalized output spectral amplitude as
\begin{equation}
u_B(\omega)=\tilde{\xi}(\omega)t_B(\omega).
\label{eq:output_spectral_amplitude}
\end{equation}
The probability of finding one photon in the pulse output is then
\begin{equation}
p_1=\int_{-\infty}^{\infty}d\omega\,
|u_B(\omega)|^2
=
\int_{-\infty}^{\infty}d\omega\,
|t_B(\omega)|^2|\tilde{\xi}(\omega)|^2 .
\label{poutFreq}
\end{equation}
These expressions are valid for an arbitrary incident single-photon pulse and show that the magnetic field modifies both the pulse-output probability and the output spectral amplitude through the joint variation of $\Delta_g$ and $\delta$. 

\subsection{Output-state QFI decomposition and measurement attainability}
\label{sec:information_decomposition}
After tracing over the three-level system and the environment, we consider the asymptotic reduced output state of the pulse. Since the control field remains on during the scattering process, any residual population in $\ket{s}$ is coherently coupled back through the lossy excited state, so that $\psi_e(t), \psi_s(t) \rightarrow 0$ as $t \rightarrow \infty$.  The reduced output state can therefore be written as
\begin{equation}
\rho_B = p_0|0\rangle\langle0|+p_1|\psi_B\rangle\langle\psi_B|,
\end{equation}
where
\begin{equation}
|\psi_B\rangle={|\psi_g^P(\infty)\rangle}/{\sqrt{p_1}},
\end{equation}
and $p_0=1-p_1$. This structure leads naturally to a decomposition of the QFI for estimating the magnetic field:
\begin{equation}
\mathcal{Q}(\rho_B)=\frac{
(\partial_B p_1)^2}{p_1[1-p_1]}+p_1 \mathcal{Q}(\ket{\psi_B}).
\label{CQFI}
\end{equation}
The two terms in Eq.~\eqref{CQFI} are often referred to as the classical and quantum contributions to the QFI, respectively. More precisely, the first is the eigenvalue contribution associated with the vacuum-versus-one-photon probabilities, while the second is the QFI of the single-photon output state weighted by its occurrence probability. For the pure single-photon output state, its QFI is~\cite{liu2014quantum,liu2019quantum}
\begin{equation}
\mathcal{Q}(\ket{\psi_B}) = 4\left[\langle \partial_B\psi_B|\partial_B\psi_B\rangle-\left|\langle \psi_B|\partial_B\psi_B\rangle\right|^2\right].
\label{QFI}
\end{equation}
Because the magnetic field changes both detunings according to Eq.~\eqref{eq:B_detuning_dependence}, derivatives with respect to $B$ must account for their simultaneous magnetic-field dependence:
\begin{equation}
\partial_B=\kappa_g\partial_{\Delta_g}+\kappa_\delta\partial_\delta .
\label{eq:B_derivative}
\end{equation}
Thus, $\partial_B$ is a single-parameter derivative: $\Delta_g$ and $\delta$ vary together with $B$ and are not treated as independently estimated parameters. The same derivative is used when evaluating any classical Fisher information (CFI). Using Eq.~\eqref{CQFI} and the output spectral amplitude $u_B(\omega)$ defined in Eq.~\eqref{eq:output_spectral_amplitude}, the output-state QFI can be written as
\begin{align}
\mathcal{Q}(\rho_B)=\frac{\left(\partial_B p_1\right)^2}{p_1\left(1-p_1\right)}+4\left[\int d\omega\,\left|\partial_B u_B(\omega)\right|^2-\frac{\left|\int d\omega\,u_B^*(\omega)\partial_B u_B(\omega)\right|^2}{p_1}\right].
\label{eq:qfi_frequency_domain}
\end{align}
Assuming that the incident spectral amplitude $\tilde{\xi}(\omega)$, the coupling rates, and the control Rabi frequency are independent of the magnetic field, the field derivative of the output spectral amplitude reduces to 
\begin{equation}
\partial_B u_B(\omega) =-\Gamma_g\tilde{\xi}(\omega)\partial_B\chi(\omega;B).
\label{eq:output_amplitude_derivative}
\end{equation}
Using
\begin{equation}
D(\omega;B)=\left[\frac{\Gamma_g+\Gamma_\perp}{2}-i\bigl(\omega+\Delta_g\bigr) \right] \bigl(\omega+\delta\bigr)+\frac{i|\Omega_c|^2}{4},
\end{equation}
we obtain
\begin{equation}
\partial_B\chi(\omega;B)=\frac{i}{D^2(\omega;B)} \left[ \kappa_g \bigl(\omega+\delta\bigr)^2+\frac{\kappa_\delta|\Omega_c|^2}{4}\right].
\label{eq:chi_field_derivative}
\end{equation}

The QFI quantifies the total magnetic-field information encoded in the output state, but does not determine how much of this information can be accessed by a specific measurement. To further resolve the quantum contribution in Eq.~\eqref{CQFI}, we decompose the single-photon output state into its normalized spectral intensity and spectral phase. We define
\begin{equation}
f_B(\omega)=\frac{|u_B(\omega)|^2}{p_1},\qquad
\int_{-\infty}^{\infty}d\omega\,f_B(\omega)=1,
\label{eq:normalized_output_spectrum}
\end{equation}
so that, on the support of $f_B(\omega)$,
\begin{equation}
u_B(\omega)=\sqrt{p_1 f_B(\omega)}\,e^{i\phi_B(\omega)},\qquad
\phi_B(\omega)=\arg u_B(\omega).
\label{eq:output_amplitude_phase_decomposition}
\end{equation}
Here, $f_B(\omega)$ is the normalized spectral intensity of the single-photon output state, and $\phi_B(\omega)$ is its spectral phase. We use $p_B$ generically to denote the magnetic-field-dependent outcome distribution of the measurement under consideration. A photon-loss measurement that distinguishes between the vacuum and one-photon components has the CFI
\begin{equation}
\mathcal{C}(p_B)=\frac{\left(\partial_Bp_1\right)^2}{p_1(1-p_1)},
\label{eq:general_binary_CFI}
\end{equation}
where, for this measurement, $p_B=\{1-p_1,p_1\}$ is the corresponding two-outcome probability distribution. By contrast, frequency-resolved photon counting has a vacuum outcome together with a continuum of one-photon outcomes labeled by $\omega$. Its CFI is 
\begin{equation}
\mathcal{C}_{\omega}(p_B)=\frac{\left(\partial_Bp_1\right)^2}{1-p_1} + \int_{-\infty}^{\infty}d\omega\, \frac{\left[\partial_B|u_B(\omega)|^2\right]^2}{|u_B(\omega)|^2}.
\label{eq:general_frequency_CFI}
\end{equation}

Using $|u_B(\omega)|^2=p_1f_B(\omega)$ and $\int d\omega\,\partial_Bf_B(\omega)=0$, Eq.~\eqref{eq:general_frequency_CFI} can be written as
\begin{equation}
\mathcal{C}_{\omega}(p_B)=\mathcal{C}(p_B)+p_1\mathcal{C}_{\mathrm{spec}}(f_B),
\label{eq:frequency_CFI_decomposition}
\end{equation}
where
\begin{equation}
\mathcal{C}_{\mathrm{spec}}(f_B)=\int_{-\infty}^{\infty}d\omega\, \frac{\left[\partial_Bf_B(\omega)\right]^2}{f_B(\omega)}=\int_{- \infty}^{\infty}d\omega\,f_B(\omega)\left[\partial_B\ln f_B(\omega)\right]^2.
\label{eq:spectral_intensity_CFI}
\end{equation}
Thus, frequency-resolved photon counting accesses both the photon-loss information and the information encoded in the spectral intensity.

Substituting Eq.~\eqref{eq:output_amplitude_phase_decomposition} into the frequency-domain QFI in Eq.~\eqref{eq:qfi_frequency_domain} gives the exact decomposition
\begin{align}
\mathcal{Q}(\rho_B)={}&\mathcal{C}(p_B)+p_1\mathcal{C}_{\mathrm{spec}}(f_B) \nonumber\\
&+ 4p_1 \left[\int_{-\infty}^{\infty}d\omega\,f_B(\omega) \left[\partial_B\phi_B(\omega)\right]^2 \
-\left(\int_{-\infty}^{\infty}d\omega\,f_B(\omega)\partial_B\phi_B(\omega) \right)^2 \right].
\label{eq:full_information_decomposition}
\end{align}
The first term in Eq.~\eqref{eq:full_information_decomposition} is the classical contribution from photon-loss statistics, while the second and third terms decompose the quantum contribution into spectral-intensity and spectral-phase contributions, respectively. Combining Eqs.~\eqref{eq:frequency_CFI_decomposition} and \eqref{eq:full_information_decomposition}, the information not captured by frequency-resolved photon counting is
\begin{equation}
\mathcal{Q}(\rho_B)-\mathcal{C}_{\omega}(p_B) = 4p_1\, \operatorname{Var}_{f_B} \left[
\partial_B\phi_B(\omega) \right] \geq 0.
\label{eq:frequency_CFI_QFI_gap}
\end{equation}
This result follows from the known geometrical decomposition of the pure-state QFI $\mathcal{Q}(|\psi_B\rangle)$ into the spectral-intensity CFI $C_{\rm spec}(f_B)$ and a spectral-phase contribution~\cite{Facchi2010}. Consequently, frequency-resolved photon counting attains the output-state QFI if and only if $\operatorname{Var}_{f_B}\left[\partial_B\phi_B(\omega)\right]=0$ or, equivalently, if $\partial_B\phi_B(\omega)$ is independent of frequency over the support of $f_B(\omega)$. Such a frequency-independent change represents only a global phase of the single-photon output state and therefore carries no additional observable information. When the variance in Eq.~\eqref{eq:frequency_CFI_QFI_gap} is nonzero, attaining the QFI generally requires a phase-sensitive coherent mode measurement.

%The resulting hierarchy is
% \begin{equation}
% \mathcal{C}(p_B)\leq\mathcal{C}_{\omega}(p_B)\leq\mathcal{Q}(\rho_B).
% \label{eq:general_information_hierarchy}
% \end{equation}
All expressions involving $\ln f_B(\omega)$ or $\phi_B(\omega)$ are understood on the support $f_B(\omega)>0$, with isolated zeros of  $f_B(\omega)$ treated by continuity. Boundary cases $p_1=0$ or $p_1=1$, for which individual formulas may contain an apparent $0/0$, are likewise defined by the corresponding continuous limit. In the present model, the lossless limit $\Gamma_\perp=0$ corresponds to unitary scattering in the pulse modes and gives $|t_B(\omega)|=1$ for real $\omega$. Hence $p_1=1$ and the photon-loss and spectral-intensity CFIs vanish, while magnetic-field information can remain encoded in the frequency-dependent spectral phase of the single-photon output state. 

\section{Results for finite pulses}\label{sec:finite_pulses}

\subsection{Gaussian input pulse}
\label{sec:gaussian_pulse}

We now specialize the general frequency-domain solution of Sec.~\ref{sec:freq_domain} and the information decomposition of Sec.~\ref{sec:information_decomposition} to a finite Gaussian single-photon pulse. We consider the real temporal profile
\begin{equation}
\xi(t)=\frac{1}{(2\pi T^2)^{1/4}}\exp\left(-\frac{t^2}{4T^2}\right),
\label{eq:GaussianRealPulse}
\end{equation}
which satisfies
$\int_{-\infty}^{\infty}dt\,|\xi(t)|^2=1$.
Here, $T$ is the pulse duration, specifically the rms duration of the temporal intensity profile. The corresponding spectral bandwidth scales as $T^{-1}$. The broadband Markovian description of Sec.~\ref{sec:interactions} therefore requires $T^{-1}\ll\omega_p$ and an approximately flat spectral density over this bandwidth. Using the Fourier convention of Eq.~\eqref{FourierConvention}, the corresponding spectral amplitude of the incident pulse is
\begin{equation}
\tilde{\xi}(\omega)=\left(\frac{2T^2}{\pi}\right)^{1/4}e^{-T^2\omega^2}.
\label{eq:GaussianSpectrum}
\end{equation}

Since the incident spectral amplitude $\tilde{\xi}(\omega)$ is independent of the magnetic field, all magnetic-field dependence of the output spectral amplitude $u_B(\omega)$ enters through $t_B(\omega)$, and hence through the response function $\chi(\omega;B)$. Because $\Delta_g$ and $\delta$ vary together with $B$ according to Eq.~\eqref{eq:B_detuning_dependence}, all field derivatives below are evaluated using Eq.~\eqref{eq:B_derivative}. Consequently,
\begin{equation}
\partial_Bu_B(\omega) = -\Gamma_g \left(\frac{2T^2}{\pi}\right)^{1/4} e^{-T^2\omega^2} \partial_B\chi(\omega;B),
\label{eq:GaussianOutputDerivative}
\end{equation}
where $\partial_B\chi(\omega;B)$ is given by Eq.~\eqref{eq:chi_field_derivative}. The derivative of the pulse-output probability is therefore 
\begin{align}
\partial_Bp_1=2\sqrt{\frac{2T^2}{\pi}}\, \operatorname{Re}\int_{-\infty}^{\infty}d\omega\,e^{-2T^2\omega^2}t_B^*(\omega)\partial_Bt_B(\omega).
\label{eq:GaussianPoutDerivative}
\end{align}
Substituting Eq.~\eqref{eq:GaussianSpectrum} into Eq.~\eqref{eq:qfi_frequency_domain} gives 
\begin{align}
\mathcal{Q}(\rho_B)={}&\frac{\left(\partial_Bp_1\right)^2}{p_1(1-p_1)}+4\sqrt{\frac{2T^2}{\pi}}\int_{-\infty}^{\infty} d\omega\, e^{-2T^2\omega^2}\left|\partial_Bt_B(\omega)\right|^2 \nonumber\\
&-\frac{8T^2}{\pi p_1}\left| \int_{-\infty}^{\infty}d\omega\, e^{-2T^2\omega^2}t_B^*(\omega)\partial_Bt_B(\omega)\right|^2 .
\label{eq:GaussianQFI}
\end{align}
The first term is the photon-loss CFI obtained from Eq.~\eqref{eq:general_binary_CFI}, using $\partial_B p_1$ from Eq.~\eqref{eq:GaussianPoutDerivative}, while the remaining two terms together give the quantum contribution $p_1\mathcal{Q}(\ket{\psi_B})$.

Substituting Eq.~\eqref{eq:GaussianSpectrum} into the general CFI for frequency-resolved photon counting in Eq.~\eqref{eq:general_frequency_CFI} gives 
\begin{align}
\mathcal{C}_{\omega}(p_B)=\frac{\left(\partial_Bp_1\right)^2}{1-p_1}+4\sqrt{\frac{2T^2}{\pi}} \int_{-\infty}^{\infty}d\omega\,e^{-2T^2\omega^2} \frac{\left\{\operatorname{Re}\!\left[t_B^*(\omega)\partial_Bt_B(\omega)\right]\right\}^2} {|t_B(\omega)|^2}.
\label{eq:GaussianFrequencyCFI}
\end{align}

Equations~\eqref{eq:GaussianQFI} and \eqref{eq:GaussianFrequencyCFI}, together with the decomposition in Sec.~\ref{sec:information_decomposition}, are used below to compare the total QFI with the photon-loss CFI and the CFI for frequency-resolved photon counting. For the numerical calculations, all frequencies are expressed in units of $\Gamma_g$. We introduce the dimensionless magnetic-field displacement and relative Zeeman sensitivity
\begin{equation}
x_B={\kappa_\delta(B-B_0)}/{\Gamma_g},\qquad 
r_Z={\kappa_g}/{\kappa_\delta},
\label{eq:dimensionless_field}
\end{equation}
so that
\begin{equation}
{\delta}/{\Gamma_g}={\delta_0}/{\Gamma_g}+x_B,\qquad
{\Delta_g}/{\Gamma_g}={\Delta_{g,0}}/{\Gamma_g}+r_Zx_B.
\label{eq:dimensionless_trajectory}
\end{equation}

In this parametrization, the two-photon resonance condition
$\delta=0$ occurs at
\begin{equation}
	x_B^{\rm EIT}=-{\delta_0}/{\Gamma_g}.
	\label{eq:xB_EIT}
\end{equation}
At this point the central spectral component $\omega=0$ is transparent,
$\chi(0;B)=0$ and hence $t_B(0)=1$. Thus, changing $\delta_0$
shifts the EIT point directly along the $x_B$ axis.

The QFI and CFI values shown below are calculated with respect to $x_B$. The parametrization in Eq.~\eqref{eq:dimensionless_field} assumes $\kappa_\delta\neq0$; when $\kappa_\delta=0$, $B$ should be used directly as the estimation parameter. The corresponding QFI and CFI values with respect to the physical magnetic field $B$ are obtained from $\mathcal{F}^{(B)}=\left({\kappa_\delta}/{\Gamma_g}\right)^2\mathcal{F}^{(x_B)}$,
where $\mathcal{F}$ denotes either QFI or CFI. All information values below are therefore per incident photon. For $N$ independent repetitions, the corresponding Cram\'er-Rao bound for the magnetic-field uncertainty is
\begin{equation}
\Delta B\geq \frac{1}{\sqrt{N\mathcal{F}^{(B)}}}
=\frac{\Gamma_g}{|\kappa_\delta|}\frac{1}{\sqrt{N\mathcal{F}^{(x_B)}}}.
\label{eq:magnetic_CR_bound}
\end{equation}

For the numerical results we use the $^{40}\mathrm{Ca}^{+}$ Zeeman configuration introduced in Sec.~\ref{sec:implementation}. Using the LS-coupling Land\'{e} factors for the selected states gives
\begin{equation}
r_Z=\frac{\kappa_g}{\kappa_\delta}=\frac{25}{33}\simeq0.76.
\label{eq:Ca_rZ_results}
\end{equation}
Unless stated otherwise, we take $\delta_0=\Delta_{g,0}=0$ and $\Omega_c/\Gamma_g=1$. The numerical integrations use Eqs.~\eqref{eq:GaussianQFI} and \eqref{eq:GaussianFrequencyCFI} with the detunings varied with $x_B$ according to Eq.~\eqref{eq:dimensionless_trajectory}.

\begin{figure}[ht]
    \centering
    \includegraphics[width=0.45\linewidth]{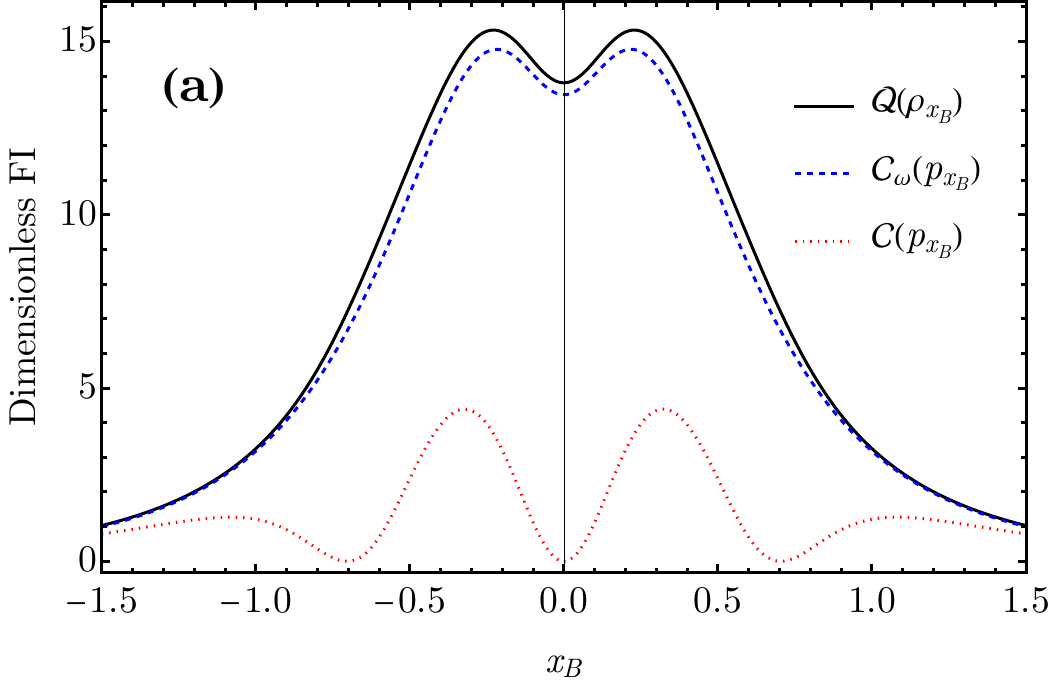}
    \includegraphics[width=0.45\linewidth]{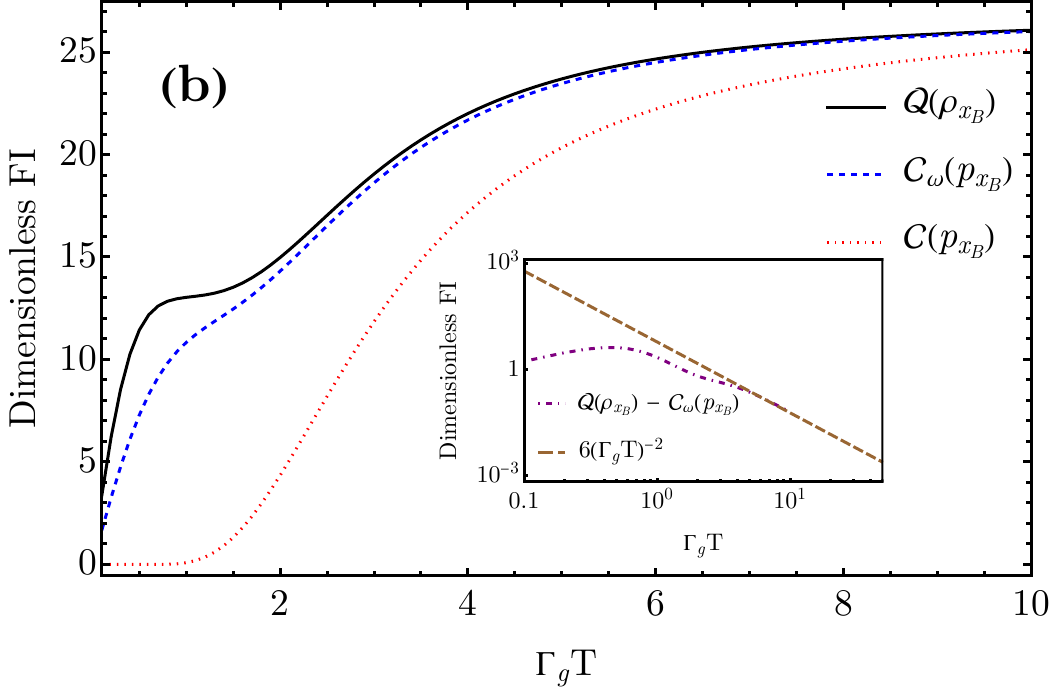}
    \caption{Magnetic-field and pulse-duration dependence for the $^{40}\mathrm{Ca}^{+}$ value $r_Z=0.76$ at critical coupling $\Gamma_\perp=\Gamma_g$, with $\Omega_c/\Gamma_g=1$ and $\delta_0=\Delta_{g,0}=0$. (a) Total output-state QFI $\mathcal{Q}(\rho_{x_B})$, frequency-resolved CFI $\mathcal{C}_\omega(p_{x_B})$, and photon-loss CFI $\mathcal{C}(p_{x_B})$ versus $x_B$ for $\Gamma_gT=2$. (b) The same quantities versus $\Gamma_gT$ at $x_B=0.3$. The inset shows the information gap $\mathcal{Q}-\mathcal{C}_\omega$ on log-log axes together with a $T^{-2}$ reference dependence; the numerical data follow the predicted asymptotic scaling at large $T$.}
    \label{fig:FI_xB_T_Ca}
\end{figure}

Figure~\ref{fig:FI_xB_T_Ca}(a) shows the total output-state QFI, the frequency-resolved CFI, and the photon-loss CFI as functions of the reduced magnetic-field displacement $x_B$ at critical coupling, $\Gamma_\perp=\Gamma_g$, for $\Gamma_gT=2$. The EIT point $x_B=0$ lies between two nearly symmetric information maxima. At this point, $\partial_{x_B}|t_B(0)|^2=0$, so the output spectral intensity at the pulse center has no first-order sensitivity to $x_B$; nevertheless, the off-center components of the finite-bandwidth pulse still carry magnetic-field information. Near these side maxima, frequency-resolved photon counting remains close to the QFI, whereas photon-loss CFI accesses only a smaller part of the available information. Figure~\ref{fig:FI_xB_T_Ca}(b) shows the corresponding dependence on pulse duration at the representative operating point $x_B=0.3$. For intermediate pulse durations, spectral resolution provides a substantial advantage over the photon-loss measurement. In the long-pulse limit, however, the spectral distribution becomes narrow enough that the photon-loss CFI also approaches the total QFI.

For a long Gaussian pulse, the normalized spectral intensity $f_B(\omega)$ becomes increasingly narrow around $\omega=0$. Provided that $t_B(\omega)$ is sufficiently smooth and nonzero at $\omega=0$, the spectral-phase contribution has the asymptotic form
\begin{equation}
\mathcal{Q}(\rho_B)-\mathcal{C}_{\omega}(p_B)
=\frac{p_1}{T^2}
\left[\left.\partial_\omega\partial_B\phi_B(\omega)\right|_{\omega=0}\right]^2
+\mathcal{O}(T^{-4}).
\label{eq:long_pulse_information_gap}
\end{equation}
Under the same regularity assumptions, the spectral-intensity contribution satisfies
$\mathcal{C}_\omega(p_B)-\mathcal{C}(p_B)=p_1\mathcal{C}_{\rm spec}(f_B)=\mathcal{O}(T^{-2})$ or faster. Consequently, both the photon-loss CFI and the frequency-resolved CFI approach the output-state QFI as $T\to\infty$. The numerical large-$T$ behavior in Fig.~\ref{fig:FI_xB_T_Ca}(b) is consistent with the $T^{-2}$ law. The derivation and regularity assumptions are given in Appendix~\ref{app:long_pulse_attainability}.

\subsection{Critical coupling, information redistribution, and scattering zeros}
\label{sec:critical_coupling}

The special role of critical coupling follows directly from the scattering amplitude $t_B(\omega)$ in Eq.~\eqref{eq:scattering_amplitude}. A zero of the atomic response should first be distinguished from a zero of the scattering amplitude: at $\omega=-\delta$, one has $\chi(\omega;B)=0$ and therefore $t_B(\omega)=1$, corresponding to transparency. For a nonzero control field, a real-frequency zero of the scattering amplitude instead requires
\begin{equation}
\Gamma_\perp=\Gamma_g,
\qquad
(\omega+\Delta_g)(\omega+\delta)=\frac{|\Omega_c|^2}{4},
\label{eq:critical_zero_conditions}
\end{equation}
and occurs at
\begin{equation}
\omega_\pm^{(0)}=-\frac{\Delta_g+\delta}{2}
\pm\frac{1}{2}\sqrt{(\Delta_g-\delta)^2+|\Omega_c|^2}.
\label{eq:transmission_zeros}
\end{equation}
Thus, $\Gamma_\perp=\Gamma_g$ is the condition that permits real-frequency zeros of the scattering amplitude, while the control-field strength and detunings determine whether those zeros overlap the incident pulse bandwidth. This is the critical-coupling condition: the incident field and the field re-emitted into the pulse modes can destructively interfere, causing the output spectral amplitude to vanish at that frequency while the photon is emitted into environmental modes.

For the central frequency of the incident pulse, $\omega=0$, the
critical-coupling zero condition can be expressed directly in terms
of the reduced magnetic-field coordinate as
\begin{equation}
	\left(
	\frac{\Delta_{g,0}}{\Gamma_g}+r_Zx_B
	\right)
	\left(
	\frac{\delta_0}{\Gamma_g}+x_B
	\right)
	=
	\frac{|\Omega_c|^2}{4\Gamma_g^2}.
	\label{eq:center_zero_general}
\end{equation}
Thus, the magnetic-field values at which a zero of the scattering amplitude crosses $\omega=0$ depend on both initial detunings and on the relative Zeeman sensitivity $r_Z$.

\begin{figure}[ht]
    \centering
    \includegraphics[width=0.45\linewidth]{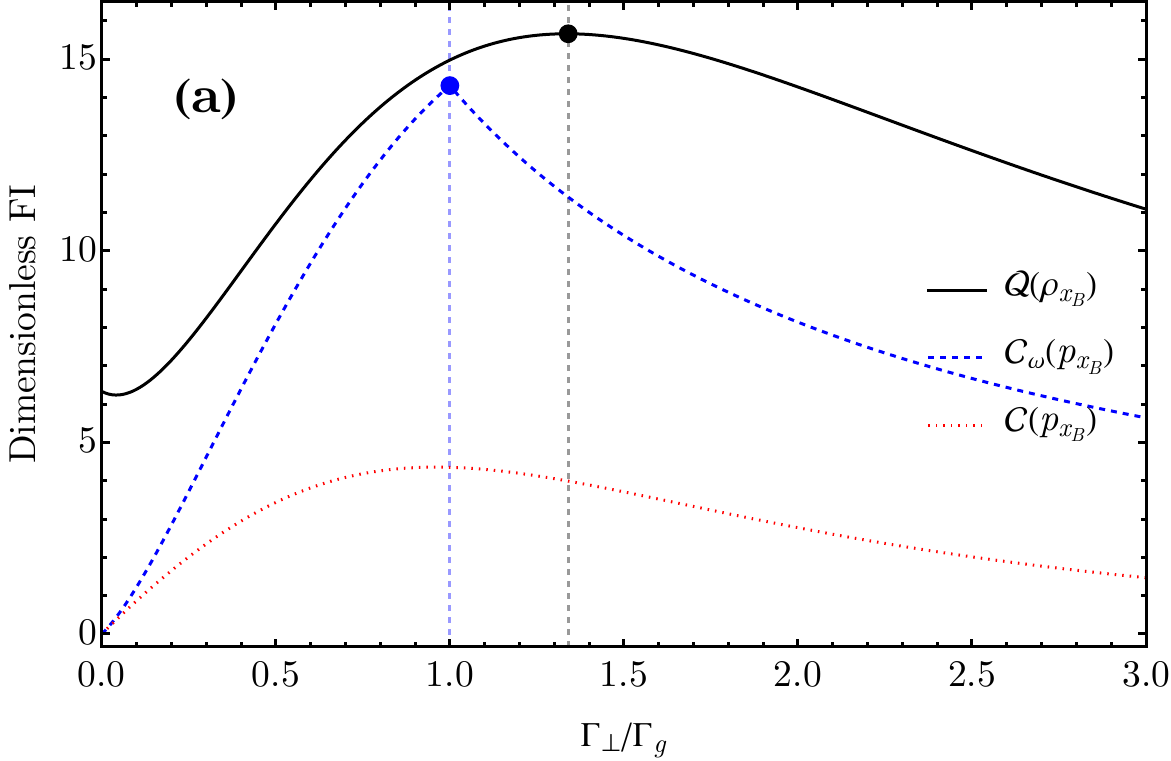}
    \includegraphics[width=0.45\linewidth]{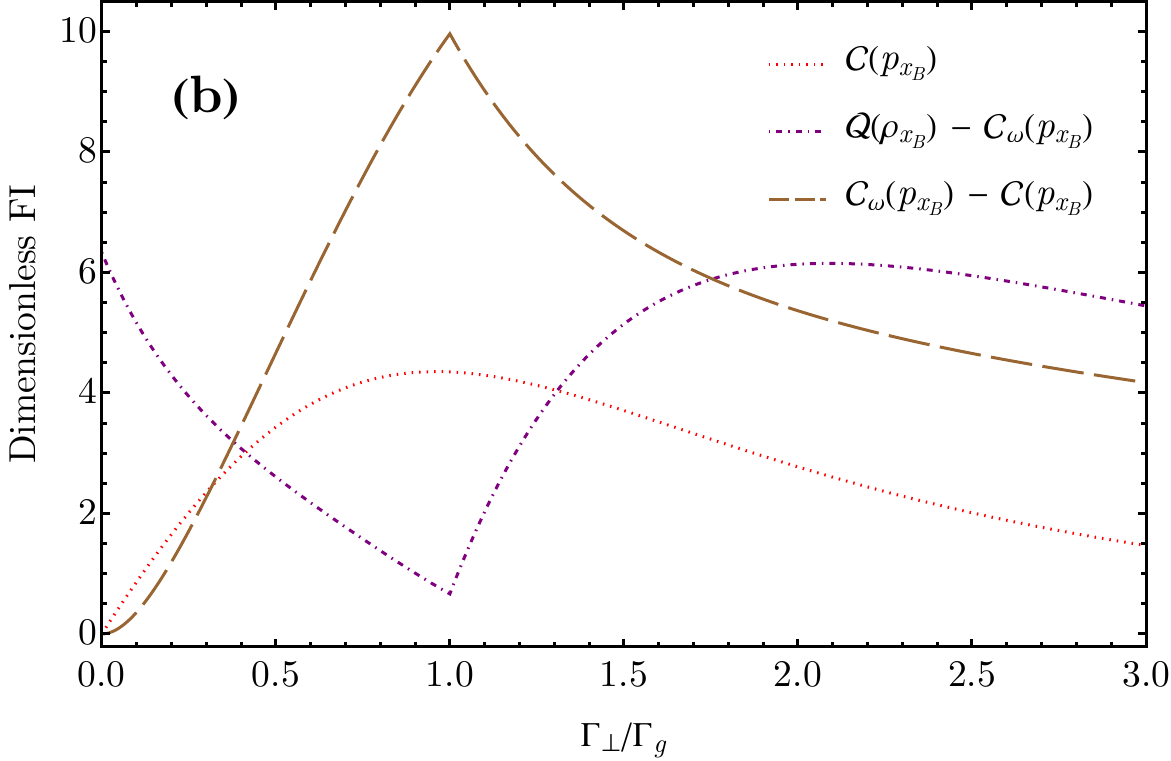}
    \caption{Information redistribution as a function of the normalized environmental coupling rate $\Gamma_\perp/\Gamma_g$ for $r_Z=0.76$, $x_B=0.3$, $\Gamma_gT=2$, $\Omega_c/\Gamma_g=1$, and $\delta_0=\Delta_{g,0}=0$. (a) Total QFI $\mathcal{Q}$, frequency-resolved CFI $\mathcal{C}_\omega$, and photon-loss CFI $\mathcal{C}$. The blue vertical line marks critical coupling, $\Gamma_\perp/\Gamma_g=1$, while the second vertical line marks the QFI maximum at $\Gamma_\perp/\Gamma_g\simeq1.34$. (b) Exact decomposition into photon-loss information $\mathcal{C}$, additional spectral-intensity information $\mathcal{C}_\omega-\mathcal{C}$, and residual spectral-phase information $\mathcal{Q}-\mathcal{C}_\omega$.}
    \label{fig:GammaDecomposition}
\end{figure}

The information redistribution associated with this critical-coupling condition is shown explicitly in Fig.~\ref{fig:GammaDecomposition}. At $\Gamma_\perp=0$, there is no coupling to the environment modes, and the scattering amplitude satisfies $|t_B(\omega)|=1$, so both $\mathcal{C}$ and $\mathcal{C}_\omega-\mathcal{C}$ vanish and the available information is encoded entirely in the spectral phase. As $\Gamma_\perp$ is increased, spectral-intensity information develops a sharp cusp and reaches its
maximum at the critical-coupling point $\Gamma_\perp/\Gamma_g=1$. The corresponding cusp remains finite because the Fisher-information contribution at a zero of the scattering amplitude is defined by its continuous limiting value and does not represent a physical divergence. At the same point, the residual spectral-phase contribution $\mathcal{Q}-\mathcal{C}_\omega$ is strongly suppressed. For the parameters of Fig.~\ref{fig:GammaDecomposition}, critical coupling maximizes both the spectral-intensity contribution $\mathcal{C}_\omega-\mathcal{C}$ and the frequency-resolved CFI $\mathcal{C}_\omega$, whereas the total QFI reaches its maximum at $\Gamma_\perp/\Gamma_g\simeq 1.34$. 

\begin{figure}[ht]
    \centering
    \includegraphics[width=0.45\linewidth]{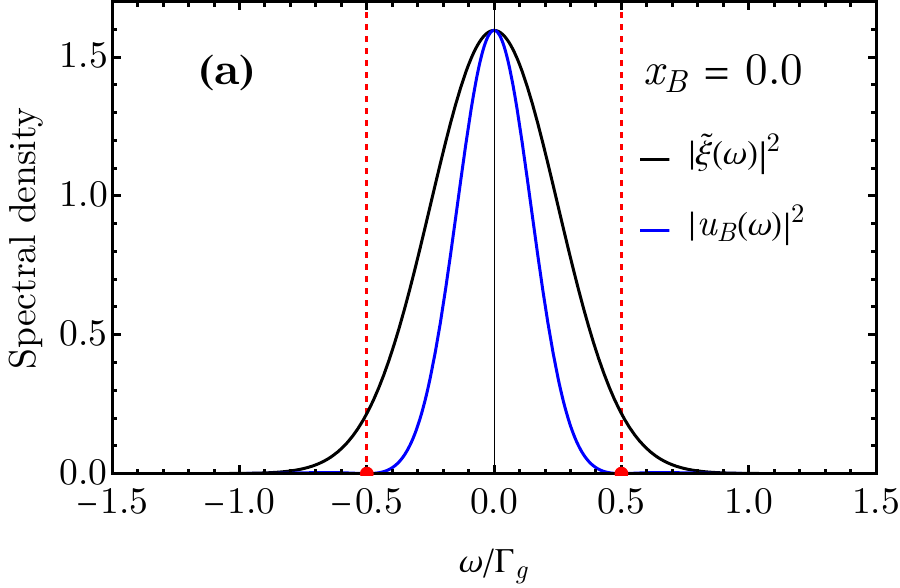}
    \includegraphics[width=0.45\linewidth]{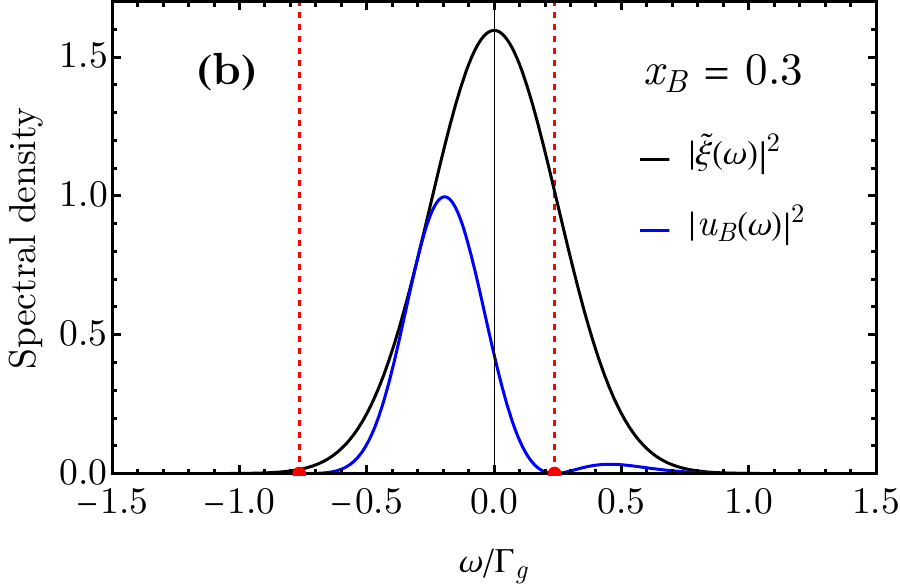}
    \\
    \includegraphics[width=0.45\linewidth]{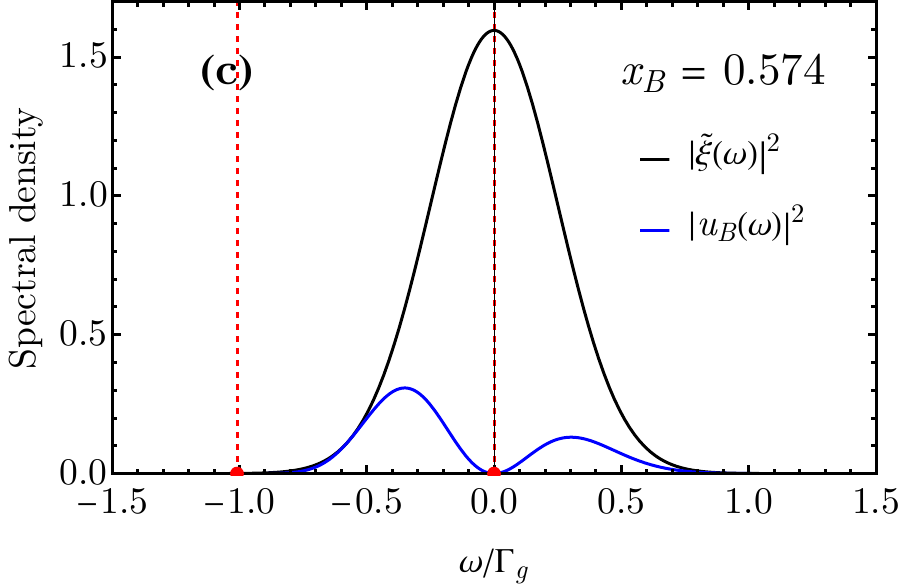}
    \includegraphics[width=0.45\linewidth]{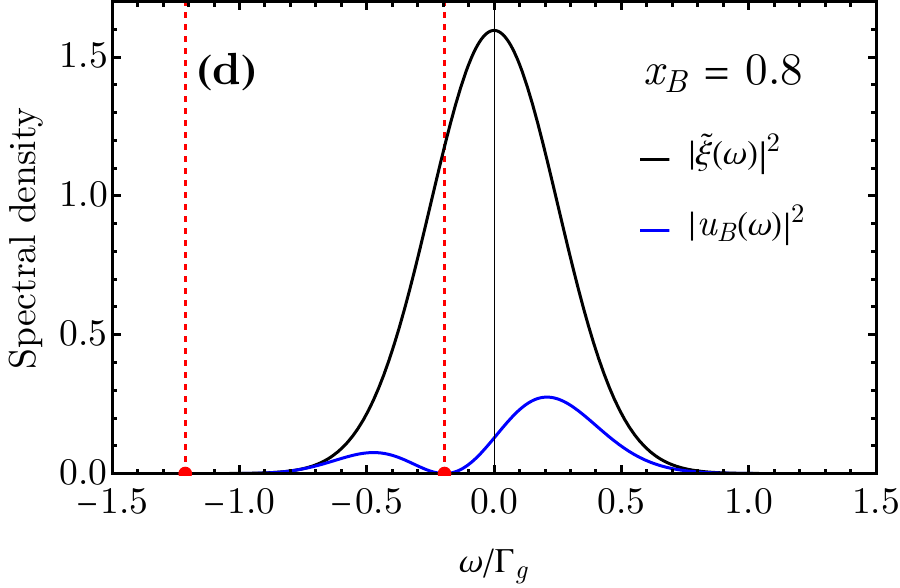}
    \caption{Incident spectral density $|\tilde\xi(\omega)|^2$ and output spectral density $|u_B(\omega)|^2$ at critical coupling $\Gamma_\perp=\Gamma_g$, for $r_Z=0.76$, $\Gamma_gT=2$, $\Omega_c/\Gamma_g=1$, and $\delta_0=\Delta_{g,0}=0$. The panels correspond to (a) $x_B=0$, (b) $x_B=0.3$, (c) $x_B=0.574$, and (d) $x_B=0.8$. Vertical dashed lines mark the analytical scattering-zero positions $\omega_\pm^{(0)}$. In panel (c), one zero lies at the pulse center.}
    \label{fig:SpectralZeros}
\end{figure}

The spectral mechanism is illustrated in Fig.~\ref{fig:SpectralZeros}. At critical coupling the zeros $\omega_\pm^{(0)}$ move continuously as the magnetic field changes. For the resonant case $\delta_0=\Delta_{g,0}=0$, Eq.~\eqref{eq:center_zero_general} gives
\begin{equation}
x_B^{(0)}=\pm\frac{|\Omega_c|}{2\Gamma_g\sqrt{r_Z}}.
\label{eq:center_zero_field}
\end{equation}
For the Ca$^+$ value $r_Z=25/33$ and $\Omega_c/\Gamma_g=1$, this yields $|x_B^{(0)}|\simeq0.574$. 
The two solutions in Eq.~\eqref{eq:center_zero_field} set characteristic magnetic-field scales for the two sides of the EIT point with enhanced sensitivity. The actual Fisher-information maxima in Fig. \ref{fig:FI_xB_T_Ca}(a) do not necessarily appear at these values because the Fisher information receives contributions from the full finite incident pulse bandwidth rather than only from $\omega=0$. Figure~\ref{fig:SpectralZeros} follows the positive-$x_B$ branch of this spectral evolution. 
At $x_B\simeq0.574$, one zero of the scattering amplitude reaches
$\omega=0$, producing complete suppression of the corresponding output spectral component. Moving beyond this point shifts the zero to the opposite side of the pulse center. The resulting motion of  a real-frequency zero of the scattering amplitude through the incident pulse bandwidth is the origin of the strong spectral-intensity sensitivity seen near critical coupling.

Near a simple zero, the scattering amplitude may be written locally as
\begin{equation}
t_B(\omega)\simeq e^{i\theta_B}A_B\,[\omega-\omega_j(B)],
\qquad A_B\in\mathbb{R},
\label{eq:local_zero_form}
\end{equation}
after extracting a frequency-independent phase $e^{i\theta_B}$. To leading order, the field-dependent deformation of the output mode is therefore a real spectral-amplitude deformation plus a global phase. Frequency-resolved photon counting captures the former, whereas only higher-order frequency-dependent phases contribute to $\mathcal{Q}-\mathcal{C}_\omega$. This local form explains why the residual spectral-phase contribution can become small when a zero of the scattering amplitude lies within
the incident pulse bandwidth. A more detailed derivation is given in Appendix~\ref{app:critical_coupling_attainability}.

\begin{figure}[ht]
    \centering
    \includegraphics[width=0.7\linewidth]{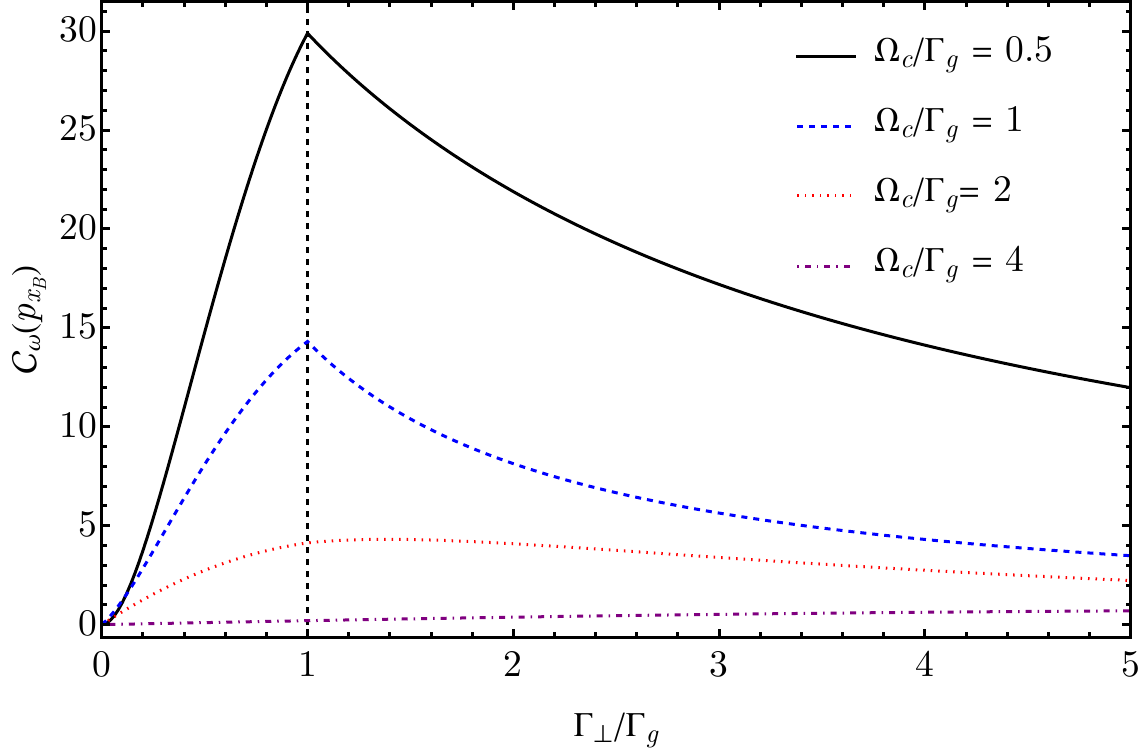}
    \caption{Frequency-resolved CFI $\mathcal{C}_\omega(p_{x_B})$ versus $\Gamma_\perp/\Gamma_g$ for $\Omega_c/\Gamma_g=0.5,1,2,4$, using $r_Z=0.76$, $x_B=0.3$, $\Gamma_gT=2$, and $\delta_0=\Delta_{g,0}=0$. The vertical dashed line marks critical coupling, $\Gamma_\perp=\Gamma_g$. The critical feature is pronounced for weak and moderate control and becomes suppressed when the scattering zeros move away from the incident pulse bandwidth.}
    \label{fig:ControlScan}
\end{figure}

Finally, Fig.~\ref{fig:ControlScan} shows that critical coupling alone does not guarantee a large frequency-resolved CFI. For $\Omega_c/\Gamma_g=0.5$ and $1$, a pronounced cusp occurs at $\Gamma_\perp/\Gamma_g=1$. As the control field is increased, the zeros separate and can move outside the incident pulse bandwidth, so that the cusp at critical coupling weakens and the global maximum of $\mathcal{C}_\omega$ shifts away from $\Gamma_\perp/\Gamma_g=1$. The control field therefore provides a direct means of tuning the positions of these zeros relative to the incident pulse bandwidth.

In summary, the numerical results reveal two distinct roles of environmental coupling. It can increase the total magnetic-field information over part of parameter space, producing a nonmonotonic QFI, and it can redistribute information between spectral phase and spectral intensity. The latter redistribution is especially strong at critical coupling, where real-frequency zeros of the scattering amplitude are allowed. For the Ca$^+$ example considered here, frequency-resolved photon counting is nearly quantum optimal over a broad parameter regime, while the exact location of the maximum total QFI remains parameter dependent.

\section{Experimental considerations and imperfections} \label{sec:implementation}

The approximations $\Gamma_s=\gamma_{gs}=0$ introduced in Sec.~\ref{sec:model} are valid when
\begin{equation}
\Gamma_s\int_{t_0}^{t_f}dt\,|\psi_e(t)|^2\ll1,\qquad
\gamma_{gs}T_{\rm int}\ll1,
\end{equation}
where $T_{\rm int}$ is the interval over which the lower-state coherence remains appreciable. Resolving the EIT feature additionally requires, up to convention-dependent numerical factors,
\begin{equation}
\gamma_{gs}\ll\Delta_{\rm EIT} \sim {|\Omega_c|^2}/{\Gamma_{\rm tot}}, \qquad
\Gamma_{\rm tot}=\Gamma_g+\Gamma_\perp .
\end{equation}

The constant-control approximation requires the nearly flat part of the control pulse to cover the entire scattering interval. The weak-probe regime with a strong classical control field has been realized in ion-based EIT experiments~\cite{Albert2018transient,Slodicka2010}. A possible implementation is a trapped $^{40}\mathrm{Ca}^+$ ion. For instance, one may choose
\begin{equation}
\ket{g}=\ket{4S_{1/2},m_g=-1/2},\qquad
\ket{e}=\ket{4P_{3/2},m_e=+1/2},\qquad
\ket{s}=\ket{3D_{3/2},m_s=+3/2}.
\label{Ca_levels}
\end{equation}
For propagation parallel to the quantization axis, the single-photon pulse may be chosen $\sigma^{+}$ polarized, with $m_e-m_g=+1$, while the classical control field may be chosen $\sigma^{-}$ polarized, with $m_e-m_s=-1$. The opposite choice, with all magnetic quantum numbers reversed and the two circular polarizations interchanged, is equally possible.

Using the LS-coupling Land\'{e} factors $\mathfrak{g}_g=2$, $\mathfrak{g}_e=4/3$, and $\mathfrak{g}_s=4/5$ for the states in Eq.~\eqref{Ca_levels}, one obtains
\begin{equation}
\kappa_g=-\frac{5}{3}\frac{\mu_B}{\hbar},\qquad
\kappa_\delta=-\frac{11}{5}\frac{\mu_B}{\hbar},\qquad
r_Z=\frac{25}{33}\simeq0.76,
\label{eq:Ca_Zeeman_slopes}
\end{equation}
corresponding to $\kappa_g/(2\pi)\simeq-2.33~\mathrm{MHz/G}$ and $\kappa_\delta/(2\pi)\simeq-3.08~\mathrm{MHz/G}$ at this level of approximation. These are the values used in the numerical results.

For this implementation, the reported total branching fraction for $4P_{3/2}\rightarrow3D_{3/2}$ is $0.00661(4)$~\cite{Gerritsma2008}. For the selected transition $m_e=+1/2\rightarrow m_s=+3/2$, the squared angular coefficient contributes a factor $2/5$, giving a branching fraction for the selected transition of approximately $2.64\times10^{-3}$ of the total $4P_{3/2}$ decay. This supports neglecting $\Gamma_s$ to leading order. The remaining $D_{3/2}$ Zeeman branches, as well as other radiative channels that are not coherently coupled back into the effective $\Lambda$ system, contribute to the environment modes. Moreover, the long lifetime of the metastable $3D_{3/2}$ state makes its population decay negligible over typical optical-pulse timescales~\cite{Kreuter2005}.

It is important that $\Gamma_g$ in the present model is the coupling rate to the pulse modes; it should not be identified with the total spontaneous-emission rate from $\ket{e}$ into the complete $4S_{1/2}$ manifold. Other spatial, polarization, and Zeeman modes are included in the environment modes and contribute to $\Gamma_\perp$ in the idealized two-continuum reduction. Accordingly, the critical-coupling condition $\Gamma_\perp=\Gamma_g$ corresponds to equal branching between the pulse and environment modes, with the branching fraction into the pulse modes given by $\beta=\Gamma_g/(\Gamma_g+\Gamma_\perp)=1/2$ within this reduced model.  

The critical-coupling condition is not fixed by the bare atomic level structure, but may be approached by enhancing the coupling to the pulse modes with an engineered optical interface or cavity~\cite{Steiner2014,Khanbekyan2017}. Representing a cavity by an effective constant rate $\Gamma_g$ requires its response to be approximately flat over the incident pulse bandwidth; outside this effective-reservoir regime, the cavity mode should be included as an explicit dynamical degree of freedom. Additional environment modes contribute to the effective $\Gamma_\perp$ and therefore modify the parameters required to reach the
critical-coupling condition, while dephasing and experimental fluctuations smooth the ideal critical-coupling feature.

The frequency-resolved CFI derived above assumes unit detection efficiency and arbitrarily fine spectral resolution. For a field-independent and frequency-independent detection efficiency $\eta$, and neglecting dark counts, the no-click probability becomes $1-\eta p_1$, and the measurable CFI is
\begin{equation}
\mathcal{C}_{\omega}^{(\eta)}=\frac{\eta^2\left(\partial_B p_1\right)^2}{1-\eta p_1}+\eta\int_{-\infty}^{\infty} d\omega \, \frac{\left[\partial_B|u_B(\omega)|^2\right]^2} {|u_B(\omega)|^2}.
\label{eq:finite_efficiency_CFI}
\end{equation}
A no-click event therefore combines the vacuum component of the output state with an undetected output photon. Finite spectral resolution further averages the output intensity over frequency bins. When the detector resolution $\delta\omega_{\rm det}$ is broader than the characteristic spectral width associated with a zero of the scattering amplitude, the corresponding CFI cusp is rounded and its magnitude is reduced. Therefore, the ideal frequency-resolved CFI provides an upper benchmark for realistic frequency-resolved photon counting.

\section{Discussion and conclusions}

We have investigated pulsed single-photon magnetometry with a Zeeman-sensitive $\Lambda$-type three-level system. The magnetic field modifies both the one-photon and two-photon detunings and is encoded in both photon-loss statistics and the spectral properties of the single-photon output state. For an arbitrary incident single-photon pulse, we derived the asymptotic reduced output state and decomposed its QFI into classical, spectral-intensity, and spectral-phase contributions.

Three main physical conclusions emerge from this analysis. First, environmental coupling is not merely detrimental. Increasing $\Gamma_\perp$ strengthens the coupling to environmental modes and therefore increases the decay rate into these modes. At the same time, it changes the interference structure of the scattering process and can enhance the magnetic-field dependence of the output state. As a result, the total output-state QFI can increase over a finite range of environmental coupling before eventually decreasing at large $\Gamma_\perp$ as the scattering response
becomes broadened and less sensitive to the magnetic field.

Second, the critical coupling $\Gamma_\perp=\Gamma_g$ has a special role because it permits real-frequency zeros of the scattering amplitude. When such a zero lies within the incident pulse bandwidth, a small change in magnetic field moves the zero across the spectrum and produces a strong redistribution of spectral intensity. In this regime, information that is encoded predominantly in the spectral phase in the lossless system becomes accessible through directly measurable spectral-intensity variations.

Third, frequency-resolved photon counting becomes nearly optimal when the residual spectral-phase contribution is small. This measurement captures both the photon-loss information and the information encoded in the spectral intensity, and near the critical-coupling regime it can recover almost the entire output-state QFI without requiring a phase-sensitive coherent measurement. For the Ca$^+$ operating point studied here, the frequency-resolved CFI is maximal at critical coupling, while the total QFI reaches its maximum at a slightly larger value $\Gamma_\perp/\Gamma_g\simeq1.34$. Thus, critical coupling is not a universal condition for maximizing the total QFI, even though it maximizes the frequency-resolved CFI at the operating point considered here.

For sufficiently long Gaussian pulses and a smooth, nonzero scattering amplitude at the pulse center, the additional spectral-intensity contribution beyond the photon-loss measurement and the residual spectral-phase information gap both decrease as $T^{-2}$ or faster. Consequently, both the photon-loss CFI and the frequency-resolved CFI approach the output-state QFI in the narrowband limit.

A possible implementation is provided by a trapped-ion $\Lambda$ system combined with an engineered optical interface. Finite spectral resolution, detector inefficiency, lower-state dephasing, and detuning fluctuations will reduce the attainable information and smooth the ideal critical-coupling feature. 
Nevertheless, the central mechanism identified here follows directly from the scattering structure: environmental coupling can enhance the magnetic-field response, critical coupling can permit real-frequency zeros of the scattering amplitude whose positions vary with the magnetic field, and frequency-resolved photon counting can convert these features into nearly quantum-optimal magnetometric information. Pulsed single-photon scattering with engineered light-matter coupling and spectral resolution therefore provides a practical route to quantum magnetometry with multilevel systems.

%\ack{E.D. and M.K. acknowledge funding by the Higher Education and Science Committee of Armenia under Grant No. 22IRF-06. }

%\funding{This work was supported by the Higher Education and Science Committee (HESC) of Armenia under Grant No. 22IRF-06.}

%\data{All numerical data supporting the findings of this study can be reproduced from the equations and parameter values given in the article. The numerical data underlying the figures are available from the authors upon reasonable request.}

\appendix
\section{Details of the time-domain scattering solution}
\label{app:time_domain_solution}
The internal amplitudes of the three-level system obey the coupled equations
\begin{align}\label{dpsie}
\dot{\psi}_e(t)&=-\left[\frac{\Gamma_g+\Gamma_\perp}{2}-i\Delta_g\right]\psi_e(t)-i\frac{\Omega_c}{2}\psi_s(t)-\sqrt{\Gamma_g}\,\xi(t),\\ \label{dpsis}
\dot{\psi}_s(t)&=i\delta\psi_s(t)-i\frac{\Omega^*_c}{2}\psi_e(t) .
\end{align}
The magnetic field affects the two coupled amplitudes through $\Delta_g$ and $\delta$. The equations can be written in matrix form as
\begin{equation}
\dot{\mathbf{v}}(t)=M\mathbf{v}(t)+\mathbf{d}(t),
\end{equation}
where
\begin{equation}
\mathbf{v}(t)=
\begin{pmatrix}
\psi_e(t)\\
\psi_s(t)
\end{pmatrix},
\label{eq:appendix_psi}
\end{equation}
and
\begin{equation}
M=
\begin{pmatrix}
-\dfrac{\Gamma_g+\Gamma_\perp}{2}+i\Delta_g&-i\Omega_c/2\\[6pt]
-i\Omega^*_c/2&i\delta
\end{pmatrix}.
\label{eq:appendix_M}
\end{equation}
The driving term due to the incoming single-photon pulse is
\begin{equation}
\mathbf{d}(t)=
\begin{pmatrix}
-\sqrt{\Gamma_g}\xi(t)\\
0
\end{pmatrix}.
\label{eq:appendix_d}
\end{equation}
We assume that the three-level system is initially in $|g\rangle$, so that $\psi_e(t_0)=\psi_s(t_0)=0$. The formal solution is then
\begin{equation}
\mathbf{v}(t)=\int_{t_0}^{t}dt'\,e^{M(t-t')}\mathbf{d}(t') .
\end{equation}
Substituting the definitions in Eqs.~\eqref{eq:appendix_psi}-\eqref{eq:appendix_d} into the formal solution yields Eq.~\eqref{eq:time_solution}
in the main text.

\section{Long-pulse limit of frequency-resolved photon counting}
\label{app:long_pulse_attainability}

Here, we derive the long-pulse result in Eq.~\eqref{eq:long_pulse_information_gap}.
For the Gaussian pulse, the normalized incident spectral intensity is
\begin{equation}
S_T(\omega)=\left|\tilde{\xi}(\omega)\right|^2=\sqrt{\frac{2T^2}{\pi}}\,e^{-2T^2\omega^2},\qquad
\int_{-\infty}^{\infty}d\omega\,S_T(\omega)=1.
\label{eq:appendix_Gaussian_spectrum}
\end{equation}
Its relevant moments are 
\begin{equation}\int d\omega\,S_T(\omega)\omega=0,\qquad
\int d\omega\,S_T(\omega)\omega^2=\frac{1}{4T^2},\qquad
\int d\omega\,S_T(\omega)\omega^4=\frac{3}{16T^4}.
\label{eq:appendix_Gaussian_moments}
\end{equation}

We write the normalized spectral intensity of the single-photon output state as 
\begin{equation}
f_B(\omega)=\frac{S_T(\omega)h_B(\omega)}{p_1},\qquad
h_B(\omega)=\left|1-\Gamma_g\chi(\omega;B)\right|^2,
\label{eq:appendix_output_spectrum}
\end{equation}
where
\begin{equation}
p_1=\int_{-\infty}^{\infty}d\omega\,S_T(\omega)h_B(\omega).
\label{eq:appendix_p1}
\end{equation}
We assume that $h_B(\omega)$ is sufficiently smooth near $\omega=0$ and that $h_B(0)>0,$ which excludes an exact zero of the scattering amplitude at the  central frequency of the pulse. Expanding $h_B(\omega)$ around
$\omega=0$,
\begin{equation}
h_B(\omega)=h_B(0)+h_B'(0)\omega+\frac{1}{2}h_B''(0)\omega^2 + \frac{1}{6}h'''_B(0)\omega^3+O(\omega^4).
\label{eq:appendix_response_expansion}
\end{equation}
Since the odd moments of $S_T(\omega)$ vanish, using the Gaussian
moments in Eq.~\eqref{eq:appendix_Gaussian_moments} gives
\begin{equation}
p_1=h_B(0)+\frac{h_B''(0)}{8T^2}+\mathcal{O}\!\left(T^{-4}\right).
\label{eq:appendix_p1_expansion}
\end{equation}
The first two moments of $f_B(\omega)$ are therefore 
\begin{align}
\left\langle\omega\right\rangle_{f_B}&=\frac{h_B'(0)}{4h_B(0)T^2}+\mathcal{O}\!\left(T^{-4}\right),
\label{eq:appendix_mean_frequency}\\
\left\langle\omega^2\right\rangle_{f_B}&=\frac{1} {4T^2}+\mathcal{O}\!\left(T^{-4}\right).
\label{eq:appendix_second_moment}
\end{align}
It follows that
\begin{equation}
\operatorname{Var}_{f_B}(\omega)=\frac{1}{4T^2}+\mathcal{O}\!\left(T^{-4}\right).
\label{eq:appendix_frequency_variance}
\end{equation}
Thus, $f_B(\omega)$ has a bandwidth proportional to $T^{-1}$, while its mean differs from zero only by a term of order $T^{-2}$. Since the definition of $C_{\rm spec}(f_B)$ can be written as $C_{\rm spec}(f_B) =\operatorname{Var}_{f_B}[\partial_B\ln h_B(\omega)]$, the same expansion gives
\begin{equation}
p_1 C_{\rm spec}(f_B) =\frac{p_1}{4T^2}\left[\left. \partial_\omega\partial_B\ln h_B(\omega)\right|_{\omega=0}\right]^2 +O(T^{-4}).
\label{eq:long-pulse-spectral-cfi}
\end{equation}
Equation~\eqref{eq:long-pulse-spectral-cfi}, together with $\mathcal{C}_\omega(p_B)=\mathcal{C}(p_B)+p_1 \mathcal{C}_{\rm spec}(f_B)$, shows that $\mathcal{C}_\omega(p_B) - \mathcal{C}(p_B)=O(T^{-2})$, or faster if the mixed derivative vanishes. Next, define
\begin{equation}
g_B(\omega)=\partial_B\phi_B(\omega).
\label{eq:appendix_phase_sensitivity}
\end{equation}
Under the same smoothness assumptions, its expansion around $\omega=0$ is
\begin{equation}
g_B(\omega)=g_B(0)+g_B'(0)\omega+\frac{1}{2}g_B''(0)\omega^2+\mathcal{O}(\omega^3).
\label{eq:appendix_phase_expansion}
\end{equation}
The constant term $g_B(0)$ does not contribute to the variance. Using Eq.~\eqref{eq:appendix_frequency_variance}, we obtain
\begin{align}
\operatorname{Var}_{f_B}\left[g_B(\omega)\right]&= \left[g_B'(0)\right]^2\operatorname{Var}_{f_B}(\omega)+\mathcal{O}\!\left(T^{-4}\right) \nonumber\\
&=\frac{1}{4T^2}\left[\left.\partial_\omega\partial_B\phi_B(\omega)\right|_{\omega=0} \right]^2+\mathcal{O}\!\left(T^{-4}\right).
\label{eq:appendix_phase_variance}
\end{align}
Finally, substituting Eq.~\eqref{eq:appendix_phase_variance} into the exact information-gap relation
\begin{equation}
\mathcal{Q}(\rho_B)-\mathcal{C}_{\omega}(p_B) =4p_1\operatorname{Var}_{f_B}\left[ \partial_B\phi_B(\omega)\right]
\label{eq:appendix_exact_gap}
\end{equation}
gives
\begin{equation}
\mathcal{Q}(\rho_B)-\mathcal{C}_{\omega}(p_B)=\frac{p_1}{T^2}\left[\left.\partial_\omega\partial_B\phi_B(\omega)\right|_{\omega=0}\right]^2 +\mathcal{O}\!\left(T^{-4}\right),
\end{equation}
which proves Eq.~\eqref{eq:long_pulse_information_gap}.
If $\partial_\omega\partial_B\phi_B(0)=0$, the leading contribution vanishes and the gap is of order $T^{-4}$ or smaller. If instead $h_B(0)=0$, the regular expansion above is invalid and the zero must be analyzed directly using the local form of $t_B(\omega)$.

\section{Near-optimal frequency-resolved photon counting at critical coupling}
\label{app:critical_coupling_attainability}

In this appendix, we show why frequency-resolved photon counting can closely approach the output-state QFI near critical coupling. The scattering amplitude is
\begin{equation}
t_B(\omega)=1-\Gamma_g\chi(\omega;B).
\label{eq:appendix_transmission_amplitude}
\end{equation}
Using the response function defined in the main text, this amplitude can be written as
\begin{equation}
t_B(\omega)=\frac{\dfrac{\Gamma_\perp-\Gamma_g}{2}(\omega+\delta)+iN_B(\omega)}{D_B(\omega)},
\label{eq:appendix_transmission_general}
\end{equation}
where
\begin{equation}
N_B(\omega)=\frac{|\Omega_c|^2}{4}-(\omega+\Delta_g)(\omega+\delta),
\label{eq:appendix_transmission_numerator}
\end{equation}
and
\begin{equation}
D_B(\omega)=\left[\frac{\Gamma_g+\Gamma_\perp}{2}-i\left(\omega+\Delta_g\right)
\right](\omega+\delta)+i\frac{|\Omega_c|^2}{4}.
\label{eq:appendix_transmission_denominator}
\end{equation}

At critical coupling, $\Gamma_\perp=\Gamma_g$, Eq.~\eqref{eq:appendix_transmission_general} reduces to
\begin{equation}
t_B(\omega)=i\frac{N_B(\omega)}{D_B(\omega)}.
\label{eq:appendix_critical_transmission}
\end{equation}
Since $N_B(\omega)$ is real for real $\omega$, the scattering amplitude vanishes at
\begin{equation}
\omega_{\pm}=-\frac{\Delta_g+\delta}{2}\pm\frac{1}{2}\sqrt{(\Delta_g-\delta)^2+|\Omega_c|^2}.
\label{eq:appendix_critical_zeros}
\end{equation}

For $\Omega_c\neq0$, these zeros are simple because $\left.\partial_\omega N_B(\omega)\right|_{\omega=\omega_\pm}\neq0$. The local expansion of $N_B(\omega)$ around either zero can be written as
\begin{equation}
N_B(\omega)=\left.\partial_\omega N_B(\omega)\right|_{\omega=\omega_j}(\omega-\omega_j)+\mathcal{O}\!\left[(\omega-\omega_j)^2\right] \qquad j\in\{+,-\}.
\label{eq:appendix_zero_expansion}
\end{equation}
Therefore, sufficiently close to either zero, the scattering amplitude can be approximated by its leading Taylor term,
\begin{equation}
t_B(\omega)\simeq\left.\partial_\omega t_B(\omega)\right|_{\omega=\omega_j} (\omega-\omega_j),\qquad j\in\{+,-\}.
\label{eq:local_transmission_appendix}
\end{equation}
This approximation is valid when the denominator of Eq.~\eqref{eq:appendix_critical_transmission} varies slowly over the spectral region surrounding the zero. The corresponding output spectral amplitude is
\begin{equation}
u_B(\omega)\simeq\tilde{\xi}(\omega)\left.\partial_\omega t_B(\omega) \right|_{\omega=\omega_j} (\omega-\omega_j).
\label{eq:local_output_appendix}
\end{equation}
Write the complex coefficient at the zero as $\left.\partial_\omega t_B\right|_{\omega_j}=e^{i\theta_B}A_B$, with $A_B$ real and positive. Equation~\eqref{eq:local_output_appendix} then becomes
\begin{equation}
u_B(\omega)\simeq e^{i\theta_B}A_B\,\tilde\xi(\omega)\,[\omega-\omega_j(B)].
\label{eq:appendix_local_real_form}
\end{equation}
To leading order, all frequency-dependent deformation caused by the magnetic field is therefore contained in a real amplitude: the zero shifts and redistributes spectral intensity, while $e^{i\theta_B}$ contributes only a frequency-independent global phase. The sign change across the zero produces the usual $\pi$ phase jump but does not generate a smooth frequency-dependent phase sensitivity away from the isolated zero. Any nonzero frequency dependence of $\partial_B\phi_B(\omega)$ in the neighbourhood dominated by the zero arises from higher-order terms omitted in Eq.~\eqref{eq:appendix_local_real_form}. Hence the residual spectral-phase contribution can be small in this regime. Using the exact relation
\begin{equation}
\mathcal{Q}(\rho_B)-\mathcal{C}_{\omega}(p_B)=4p_1\operatorname{Var}_{f_B} \left[\partial_B\phi_B(\omega)\right],
\end{equation}
we obtain $\mathcal{C}_{\omega}(p_B)\simeq\mathcal{Q}(\rho_B)$. 
This result explains why frequency-resolved photon counting nearly saturates the output-state QFI when at least one zero of the scattering amplitude lies within the incident pulse bandwidth. Critical coupling alone does not guarantee near saturation if the zeros lie outside this bandwidth.

\bibliographystyle{apsrev4-2}
\bibliography{refs}

\end{document}